\documentclass[12pt]{article}
\usepackage{epsfig}
\usepackage{amsfonts}
\usepackage{amsopn}
\usepackage{amsmath}
\usepackage{verbatim,amsthm}

\title
{
\vskip-50 pt
\begin{flushright}
\normalsize\rm NORDITA-2011-50
\end{flushright}
\vskip 20 pt
Toroidal $p$-branes,
 anharmonic oscillators and (hyper)elliptic solutions
}

\author{
 A. A. Zheltukhin $^{a,b,c}$\thanks{e-mail: aaz@physto.se}  \\ \\
$^a$ Kharkov Institute of Physics and Technology, \\
1, Akademicheskaya St., Kharkov, 61108, Ukraine \\  
$^b$ Physics Department, Stockholm University, AlbaNova,  \\
106 91, Stockholm, Sweden \\ 
$^c$ NORDITA,  \\
Roslagstullsbacken 23, 106 91 Stockholm, Sweden
}

\date{}

\begin{document}

\maketitle

\begin{abstract}
Exact solvability of brane equations is studied, and a 
new $U(1)\times U(1)\times\ldots \times U(1)$ invariant 
 anzats for the solution of  $p$-brane equations in $D=(2p+1)$-dimensional
 Minkowski space is proposed. The reduction of the $p$-brane Hamiltonian to the 
  Hamiltonian of $p$-dimensional relativistic anharmonic oscillator with the monomial 
  potential of the degree equal to $2p$ is revealed. For the case of degenerate p-torus
 with equal radii it is shown that the $p$-brane equations are integrable and their 
 solutions are expressed in terms of elliptic ($p=2$) or hyperelliptic ($p>2$) functions.
The solution describes contracting $p$-brane with the contraction time 
depending on $p$ and  the brane energy  density.
The toroidal brane elasticity is found to break down linear Hooke law 
as it takes place for the anharmonic elasticity of smectic liquid crystals.
\end{abstract}

\section{Introduction}

Membranes and p-branes play an important role in M/string theory \cite{M1}, 
and their quantization is one of current hot problems. 
Their solution is complicated by nonlinearity of the classical brane equations 
in contrast to the string  (p=1) case \cite{Reb}. 
The question of the membrane (p=2) and $p$-brane dynamics and its 
integrability attracts much attention (see e.g. [3-15] 
\nocite{tucker, hoppe1, BST, DHIS, FI, WHN, Z_0, WLN, BZ_0, hoppe2, Pol, UZ, hoppe6}).
P-branes belong to Hamiltonian dynamical systems for which there is 
known the notion 
of complete integrabilility in the Lioville sense. 
The notion implies the existence of
a maximal set of functions on the phase space which have zero Poisson 
brackets between themselves and with the system Hamiltonian. 
These functions are time-independent and split 
the whole phase space into hypersurfaces which are closed in the process of 
the system evolution.
A finite dimensional phase space has the dimension equal to $2n$ 
and the maximal number of such independent invariants is $n$, respectively.
 If a Hamiltonian $H$ and associated  PB's do not explicitly depend 
on time then $H$ belongs to one of the mentioned invariants, and the 
system preserves its energy coinciding with  $H$. 
For the case of compact energy levels the invariant hypersurfaces 
of completely integrable systems 
may be transformed into  two-dimensional tori associated with special 
sets of canonical pairs 
called the action and angle variables \cite{LLm}. The action variables are 
preserved integrals constructed by integration of the one-forms 
$pdq$ along the cycles of invariant tori on which the 
angle variables have the sense of angle parameters. 
The Hamiltonian equations for these angle variables turn out to be linear. 
Thus, the question of complete integrability of general 
Hamiltonian equtions may be reformulated into the equivalent problem 
of an explicit construction of the angle action variables. 
The Hamilton-Jacobi method helps to construct such variables using 
special canonical transformations which result in 
 a new Hamiltonian independent of all generalized coordinates. 
Then the integrability problem is reduced to the construction of  
complete solution of the corresponding Hamilton-Jacobi equation. 
P-branes are field systems with an infinite number of the 
degrees of freedom, because their world vectors $ x_{m}$ 
depend on continious world-volume parameters.
 Generalization of the notion of complete Lioville 
integrability for Hamiltonian nonlinear PDE's implies the 
existence of an infinite number of Poisson commuting invariants. 
The sine-Gordon, KdV and KP equations give well known examples
 of completely integrable equations in two and three dimensions, 
respectively. 
In addition to the existence of an infinite number of conservation 
laws these equations  are characterized by the presence 
of Backlund transformations and solitonic solutions which describe  
non-perturbative sectors in the space of solutions. 
The Backlund transformations permit to generate one-parametric family 
of new solutions starting from one known partial solition.
As a result, the given integrals of motion become dependent on 
the Backlund parameter, and their expansion in this parameter yields 
an infinite number of new conservation laws.
The Backlund parameter is connected with the time-independent eigenvalue 
of isospectral linear eigenvalue problem which plays a fundamental role 
in the inverse scattering method.
A key role in the proof of integrability belongs to so called 
Lax pair of some linear operators such that their Poisson bracket reproduces
 the nonlinear equation under investigation (see e.g. \cite{FT}). 
This  gives a general background 
for discussing  the $p$-brane integrability problem. 

A direct application of the inverse spectral method
 to $p$-brane equations meets essential problems, but 
 there is some progress  
in study of various particular solutions of the equations.
  However, not so much is known about such  solutions. 
While studying this problem Hoppe  proposed 
the  $U(1)$ invariant anzats for closed membranes and reformulated their 
equations in D=5 into the system of 2-dim nonlinear equations \cite{JU1}. 
 The elliptic solution of these equations, describing a family of closed 
contracting 2d tori, together with the solution corresponding 
 to a spinning 2d torus, found in \cite{TZ} give an example of 
time-dependent particular solutions.
 The static equations of  $U(1)$ invariant membranes in $D=(2N+1)$-dimensional 
Minkowski space were integrated 
and their general solution  for any $N>1$ was presented in \cite{TZ}.
 The geometric approach to the description of the invariant membranes 
in D=5 developed in \cite{ZT} revealed their connection 
with the nonlinear
 pendulum equation and a two-dimensional generalization of the 
nonlinear Abel equation.   

Here we extend this approach to branes and propose an anzats for the solution of 
p-brane equations in $D=(2p+1)$-dimensional
 Minkowski space with any integer $p>1$.  The strong  $(D,p)$-correlation
  between the space-time and brane dimensions 
 covers, in particular,  interesting cases of globally invariant 5-branes of M/string 
 theory in D=11 space-time, 
  membranes in D=5 and 3-branes in D=7.
The proposed anzats corresponds to closed compact p-branes with the global rotational 
symmetry  $U(1)\times U(1)\times\ldots \times U(1)$ (with p multipliers) of their 
p-dimensional hypersurfaces $\Sigma_{p}$.  
These hypersurfaces turn out to be isometric to flat $p$-tori with zero curvatures. 
The Hamiltonians and equations of invariant 
p-branes are constructed,  and it is shown that they describe p-dimensional anharmonic 
oscillators with the quartic potential 
for membranes in D=5, and the monomial potential of the degree $2p$ for 
the p-brane in D=2p+1. 
A characteristic feature of these Hamiltonians is the absence  of  the 
quadratic terms  which 
are contained in the Hamiltonian
 of the harmonic oscillator \footnote{Usually the notion of anharmonic 
oscillator is used in the case when the quartic  
 and higher terms in the potential energy are small in comparison with its 
quadratic terms.
 Here we use  this term  despite 
 the absence of the quadratic term in the p-brane Hamiltonians 
and do not assume the smallness of the higher monomials.}.
The p-brane equations are reformulated into the equations for an elastic 
 media with 
the symmetric stress tensor corresponding to isotropic pressure.
 For the case of degenerate p-torus 
with all equal radii these nonlinear equations are integrable
 and their solutions are expressed 
in terms of the elliptic cosine for $p=2$ or hyperelliptic functions 
for higher $p>2$. 
The solutions describe contracting $p$-branes with the contraction
 time depending on their dimension $p$ and energy density. 
The dependence of $p$ is partially localized in the Euler beta function.
We revealed that an elastic medium, associated with the $p$-toroidal branes, 
is characterized by a nonlinear elasticity law which generalizes the 
linear Hooke law. 
The brane elasticity is found to be similar to the anharmonic elasticity 
 earlier discovered in the statistical physics of 2D and 3D smectic 
liquid crystals \cite{C}, \cite{GP}, \cite{GZ}.

\section{P-brane dynamics for any $(D,p)$}

The Dirac action for a p-brane without boundaries
is defined by the integral in the dimensionless worldvolume
parameters $\xi^{\alpha}$ ($\alpha=0,\ldots,p$)
\footnote {Here the D-dimensional Minkowski space has
the signature $\eta_{mn}=(+,-,\ldots,-)$.}
\[
S=T\int \sqrt{|G|}d^{p+1}\xi, \label{1}
\]
where $G$ is the determinant of the induced metric $G_{\alpha
\beta}:=\partial_{\alpha} x_{m}\partial_{\beta} x^{m}$ and $T$ is 
the p-brane tension with the dimension $L^{-(p+1)}$, 
because $x^{m}$ has the  dimension of   length.
\label{2} After splitting of the $p$-brane world vector
$x^{m}=(x^0,x^i)=(t,\vec{x})$ and internal coordinates
$\xi^{\alpha}=(\tau,\sigma^r)$, the Euler-Lagrange equations and
$(p+1)$ primary constraints generated by $S$ take the  form
\begin{equation}\label{5}
\partial_{\tau}{\mathcal{P}}^{m}=-T\partial_r(\sqrt{|G|}G^{r\alpha}\partial_{\alpha}x^{m}), 
\ \ \
\mathcal{P}^{m}=T\sqrt{|G|}G^{\tau \beta}\partial_{\beta} x^{m},
\end{equation}
\begin{equation}
\tilde{T}_{r}:=\mathcal{P}^{m} \partial_{r} x_{m} \approx 0, \ \ \ \
\tilde{U}:=\mathcal{P}^{m}\mathcal{P}_{m}-T^{2}|\det G_{rs}| \approx  0, \label{6}
\end{equation}
where $\mathcal{P}^{m}$ is the energy-momentum density.
Then we use the orthogonal gauge simplifying  the metric $G_{\alpha\beta}$  
\begin{eqnarray} \label{7}
L\tau=x^0\equiv t, \ \ \ \ G_{\tau r}= -L(\dot{\vec{x}} \cdot \partial_r \vec{x})=0, \\
 g_{rs}:=\partial_r \vec{x} \cdot \partial_s \vec{x}, \ \ \ \
G_{\alpha\beta}=\left( \begin{array}{cc}
                       L^{2}(1- {\dot{\vec{x}}}^2)& 0    \\
                         0 & -g_{rs}
                              \end{array} \right)
                               \nonumber
\end{eqnarray}
with $\dot{\vec{x}}:=\partial_{t}\vec{x}= L^{-1}\partial_{\tau}\vec{x}$.
The solution of the constraint $\tilde{U}$ (\ref{6}) has the form
\begin{equation}
\mathcal{P}_0=\sqrt{\vec{\mathcal{P}}^2+T^{2}|g|}, \ \ \ \
g=\det(g_{rs}) \label{9}
\end{equation}
and becomes the Hamiltonian density $\mathcal{H}_0$ for  the p-brane
since $\dot{\mathcal{P}}_0=0$ in view of Eq. (\ref{5}).
Using the definition of $\mathcal{P}_0$  (\ref{5}) and
 $G^{\tau\tau}={1}/{ L^{2}(1-\dot{\vec{x}}^2)}$ 
 we find the  expression of $\mathcal{P}_0$ as a function of the p-brane 
 velocity $\dot{\vec{x}}$
\begin{equation}\label{5'}
\mathcal{P}_0:=TL\sqrt{|G|}G^{\tau\tau}=T\sqrt{\frac{|g|}{1-\dot{\vec{x}}^2}} \,\,\,.
\end{equation}
Taking into account this expression  and the definition (\ref{5}) one 
can present $\vec{\mathcal{P}}$ and its evolution equation (\ref{5}) as
\begin{equation}\label{13}
\vec{\mathcal{P}}=\mathcal{P}_0 \dot{\vec{x}}
 , \ \ \ \
\dot{\vec{\mathcal{P}}}= 
T^{2}\partial_r \left( \frac{|g|}{\mathcal{P}_0} g^{rs}\partial_s \vec{x}\right).
\end{equation}
Then Eqs. (\ref{13}) yield the second-order PDE for $\vec{x}$
\begin{equation}\label{xeqv}
 \ddot{\vec{x}}=
\frac{T}{\mathcal{P}_0}\partial_r \left( \frac{T}{\mathcal{P}_0}|g|g^{rs}\partial_s \vec{x}\right).
\end{equation}
These equations may be presented in the canonical Hamiltonian  form
\[
\dot{\vec{x}}=\{H_{0}
,\vec{x}\}, \ \ \ \ \dot{\vec{\mathcal{P}}}=\{H_{0},\vec{\mathcal{P}}\}, \ \ \ \
\{\mathcal{P}_i(\sigma), x_j(\tilde{\sigma})  \}=
\delta_{ij}\delta^{(p)}(\sigma^r-\tilde{\sigma}^r)
\]
using the integrated Hamiltonian density (\ref{9}) $\mathcal{H}_0(=\mathcal{P}_0)$
\begin{eqnarray}\label{hampbr}
H_0=
\int d^p \sigma \sqrt{\vec{\mathcal{P}}^2+T^{2}|g|}.
\end{eqnarray}
The presence of square root in (\ref{hampbr}) points out on the presence of the known 
residual symmetry preserving the orthogonal gauge (\ref{7}) 
\begin{equation}\label{diff}
\tilde{t}=t, \ \ \ \ \tilde{\sigma}^r=f^r(\sigma^s)
\end{equation}
and generated by the constraints $\tilde{T}_r$  (\ref{6}) reduced to the form
\begin{equation}\label{T}
T_r:=\vec{\mathcal{P}}\partial_r\vec{x}=0 \ \ \    \Leftrightarrow \ \ \ 
 \dot{\vec{x}} \partial_r\vec{x}=0,\ \ \ (r=1,2, \ldots, p).
\end{equation}
 The gauge  freedom associated  with the symmetry (\ref{diff}) allows to put $p$ 
additional time-independent conditions on $\vec{x}$ and its space-like derivatives. 

The above description of brane dynamics is valid for any 
 space-time and brane world-volume dimensions $(D,p)$ (with $D>p$).

\section{ $U(1)\times U(1)\times \ldots \times U(1)$ invariant p-branes}

Here we consider  p-branes evolving in $D=(2p+1)$-dimensional Minkowski 
space-time, and assume that their shape 
is invariant under the direct product  ${\cal U}:=\prod_{a=1}^{p} U_{a}(1)$. 
Each of these $U(1)$ symmetries is 
locally isomorphic to one of the  $O(2)$ subgroups of the $SO(2p)$ 
subgroup of the Euclidean rotations.   
 Thus, the p-dimensional hypersurface  $\Sigma_{p}$ of ${\cal U}$ invariant 
p-brane has the group ${\cal U}$ as its isometry with $p$ Killing vectors.
This points to the existence of a parametrization of the
 p-brane hypersurface $\Sigma_{p} $ with the metric tensor $g_{rs}$
 independent of $\sigma^r$.
 Our analysis will be restricted by the case of ${\cal U}$ 
invariant p-branes without boundaries. The invariant p-branes with
 boundaries are treated similarly
taking into account additional boundary terms.

To construct ${\cal U}$ invariant p-brane hypersurface,
we propose the following anzats for its space world vector $\vec{x}$ 
\begin{eqnarray} 
\vec{x}^T
=(q_1\cos\theta_1,q_1\sin\theta_1,q_2\cos\theta_2,q_2\sin\theta_2,
\ldots,q_p\cos\theta_p,q_p\sin\theta_p), \label{anzats}
 \\
q_a=q_a(t), \, \, \, \, \, \ \ \ \  \theta_a=\theta_a(\sigma^r), 
\, \, \, \, \, \,  (a=1,\ldots,p), \nonumber
\end{eqnarray}
where $T$ is the transposition of the column usually used for a vector components.
This space vector lies in the $2p$-dimensional Euclidean subspace of 
 $(2p+1)$-dimensional Minkowski space and automatically satisfies  the orthogonality 
 constraints (\ref{T}): $\dot{\vec{x}} \partial_r\vec{x}=0$.
The anzats (\ref{anzats}) originates from the realization 
of $2p$-dimensional vector $\vec{x}$ describing any p-brane, and is defined by $p$ pairs 
of its "polar" coordinates 
\begin{eqnarray}\label{Gen} 
\vec{x}^T(t,\sigma^{r})=
(q_1\cos\theta_1,q_1\sin\theta_1,\ldots,q_p\cos\theta_p,q_p\sin\theta_p)
\end{eqnarray}
with the coordinates $q_a, \theta_a$  depending on all the parameters 
  $(t,\sigma^{1},..., \sigma^{p})$ of the p-brane 
  world volume: $q_a=q_a(t,\sigma^{r}), \,  \theta_a=\theta_a(t,\sigma^{r})$.

Thus, the proposed anzats (\ref{anzats}) is obtained from the general  
representation (\ref{Gen})
 by excluding the time dependence for all "polar"
 angles $\theta_a=\theta_a(\sigma^r)$ as well as $\sigma^{r}$ dependence for 
 all "radial" coordinates $q_a=q_a(t)$. As a result, at any fixed moment $t$ 
the vector $\vec{x}^T$ (\ref{anzats}) is produced  from the vector 
 $\vec{x_0}^T=(q_1,0,q_2,0,\ldots,q_p,0 )$
by the transformations of the diagonal subgroup ${\cal U}\in SO(2p)$,
  parametrized by the angles $\theta_{a}$which  rotate the 
planes $x_1x_2$, $x_3x_4$ ,...,$x_{2p-1}x_{2p}$. 

So, the anzats (\ref{anzats}) describes one of the representatives 
of the family of ${\cal U}$ invariant (hyper)surfaces 
embedded in the $2p$-dimensional Euclidean space. 
Each of the members of the family has the symmetry ${\cal U}$ 
 as its inherent symmetry. The $p$-brane worldvolume 
metric $G_{\alpha \beta}$ corresponding to (\ref{anzats}) 
has the form  similar to (\ref{7}) with the non-zero components 
\begin{eqnarray}\label{metr} 
 G_{tt}=1- \dot{\bf{q}}^2, \  \  \  {\bf q}:=(q_1,..,q_p), \  \ \
g_{rs}= \sum_{a=1}^{p}q_{a}^{2}\theta_{a,r}\theta_{a,s},
\end{eqnarray}  
where $\theta_{a,r}\equiv \partial_{r}\theta_{a}, \ \  \dot{\bf{q}}\equiv \partial_{t}\bf{q}$, 
and yields the following interval $ds^2$ on $\Sigma_{p+1}$ 
\begin{equation}\label{intr}   
ds^2_{p+1}=(1-\dot{\bf{q}}^2)dt^{2} -  \sum_{a=1}^{p}q_{a}^{2}(t)d\theta_{a}d\theta_{a}.
\end{equation}
 This representation shows that the change of $\sigma^{r}$ by the new 
 coordinates $\theta_{a}(\sigma^{r})$, parametrizing $\Sigma_{p}$, 
 makes the transformed metric independent of  $\sigma^{r}$.  
 In the  $\theta_{a}$ parametrization the above-mentioned 
 Killing vector fields on $\Sigma_{p}$ take the form of the 
 derivatives  $\frac{\partial}{\partial \theta_a}$.
All that shows that  ${\cal U}$ invariant (hyper)surface (\ref{anzats}) 
has zero curvature and is isometric to a flat p-dimensional 
torus $S^1\times S^1\times \ldots \times S^1$ (with p multipliers $S^1$) at any fixed 
moment of time.

The canonical momentum components $\pi_a=\frac{\partial \mathcal{L}}{\partial \dot{q_a}}=
\vec{\mathcal{P}}\frac{\partial{\dot{\vec{x}}}}{\partial \dot{q_a}},
\  (a=1,2,.., p)$,
$\boldsymbol{\pi}:=(\pi_1,..., \pi_p)$ conjugate to the
 coordinates $\bf{q}$  (\ref{anzats}) may be presented in 
 the explicit form as
\begin{equation}\label{Imp}
\boldsymbol{\pi} = \mathcal{P}_0 \dot{\bf{q}}, \ \ \
\mathcal{P}_0 =T\sqrt{\frac{|g|}{1-\dot{\bf{q}}^2}} \ \ 
\end{equation}
after using  (\ref{13}) and the relations
\begin{equation}\label{mtrq}   
\vec{x}^2={\bf q}^2, \ \ \ 
\dot{\vec{x}}^2=\dot{\bf{q}}^2, \ \ \
\ \ \ g=det(\sum_{a=1}^{p} q_{a}^{2}\theta_{a,r}\theta_{a,s}). 
\end{equation}
Then the Hamiltonian density (\ref{9}) in the $(\bf{q},\boldsymbol{\pi})$ 
phase space takes the form
\begin{equation}\label{Hdns} 
\mathcal{H}_0=\mathcal{P}_0=\sqrt{\boldsymbol{\pi}^2+T^{2}|g|}, \ \ \
\dot{\mathcal{P}_0}=0
\end{equation}
that matches with representation (\ref{Imp}) for  $\mathcal{P}_0$ through 
the velocity $\dot{\bf{q}}$
\begin{equation}\label{T1}
\mathcal{P}_0= T\sqrt{\frac{|g|}{1- \dot{\bf{q}}^2}}.
\end{equation}
The corresponding Hamiltonian
equations are transformed into  the equations
\begin{equation} \label{hameq}
\dot{\bf{q}}=\{H_0,{\bf{q}}\}=\frac{1}{\mathcal{P}_0}{\boldsymbol{\pi}},
\ \ \ \
\dot{\boldsymbol{\pi}}=\{H_0,\boldsymbol{\pi}\},
\end{equation}
accompanied with the standard Poisson brackets and the Hamiltonian $H_0$
\begin{equation}\label{sqrham}
H_0=\int d^p\sigma \sqrt{\boldsymbol{\pi}^2+ T^{2}|g|},\ \ \
\{\pi_{a}, q_b\}=\delta_{ab}, \ \ \ \{q_{a}, q_b\}=\{\pi_{a}, \pi_b\}=0. 
\end{equation}
In the next section we study the equations of motion
 for the  coordinates of the effective vector $\bf{q}$.

\section{Equations of ${\cal U}$ invariant p-branes}

To simplify the equations of the introduced ${\cal U}$ invariant p-brane
 we choose the functions $\theta_{a}(\sigma^r)$ 
in (\ref{anzats}) to be linear. 
This can be done by additional gauge fixing 
 for the residual gauge symmetry 
 (\ref{diff}):  $\theta_{a}=\delta_{ar} \sigma^{r}$, 
 i.e. as
 \footnote{To cover the case of p-brane with windings one can 
 fix the gauge conditions as $\theta_{a}=n_{a}\delta_{ar} \sigma^{r}$,
 where $(n_{1}, n_{2},..., n_{p})$ 
 are the integer numbers corresponding to the winding numbers on the 
circles $0\leq\sigma^{r} \leq 2\pi$
  parametrized by $\sigma^{r}$.} 
\begin{equation}\label{qgauge}
\theta_{1}(\sigma^{r})=\sigma^{1}, \ \ \ \theta_{2}(\sigma^{r})=\sigma^{2}, \ 
 \ldots \  , \  \theta_{p}(\sigma^{r})=\sigma^{p}.
\end{equation}
In the gauge (\ref{qgauge}) the anzats (\ref{anzats}) and $g_{rs}$ (\ref{metr}) 
are expressed as follows
\begin{eqnarray}
\vec{x}^T(t)=(q_1\cos\sigma^1,q_1\sin\sigma^1, \ldots , q_p\cos\sigma^p,q_p\sin\sigma^p),
\label{ganz} \\
g_{rs}(t)= q_{r}^{2}(t)\delta_{rs}, \ \ \
 g=(q_{1} q_{2}...q_{p})^{2} \label{gmtr}  
\end{eqnarray}
with the diagonalized metric $g_{rs}(t)$ depending only  on the time $t$. 

As a result, the interval (\ref{intr}) and the inverse metric 
tensor  $g^{rs}$  reduced on $\Sigma_{p}$ take the form
\begin{equation}\label{gintr}   
ds_{p}^2= \sum_{r=1}^{p}q_{r}^{2}(t)(d\sigma^{r})^{2},  \ \ \ 
 g^{rs}(t)= \frac{1}{q_{r}^{2}}\delta_{rs}.
\end{equation}
 Gauge (\ref{qgauge}) clarifies that coordinates  
${\bf{q}}(t)=(q_{1},q_{2},\ldots,q_{p})$ are the 
time dependent radii ${\bf{R}}(t)=(R_{1},R_{2},\ldots,R_{p})$  
of the flat $p$-torus $\Sigma_{p}$.  
In the gauge (\ref{qgauge}) the Hamiltonian density $ \mathcal{H}_0$  
 becomes independent of the $p$-torus parameters $\sigma^{r}$ 
 and reduces to a constant C chosen to be positive 
\begin{equation}\label{gHdns} 
\mathcal{H}_0=\mathcal{P}_0=
\sqrt{\boldsymbol{\pi}^2+T^{2}\prod_{r=1}^{p}q_{r}^{2}} =C, \ \ \
\dot{\mathcal{P}_0}=\frac{\partial}{\partial\sigma^r}\mathcal{P}_0=0.
\end{equation}
 Equations (\ref{gHdns}) combined with (\ref {T1})  are equivalent to the initial
 data condition 
\begin{equation}\label{gHC} 
\mathcal{P}_0=C \ \ \ \rightarrow  \ \ \ 
T\sqrt{\frac{(q_{1}q_{2} \ldots q_{p})^2}{1- \dot{\bf{q}}^2}} = C.
\end{equation}  
Then  Eqs. (\ref{13}) for the
vector  $\vec{x}$ are simplified to the form
\begin{eqnarray}\label{xeqvg}
 \ddot{\vec{x}} - (\frac{T}{C})^2 gg^{rs} \partial_{rs} \vec{x}=0.
\end{eqnarray}
 Let us take into account the relations  following
from (\ref{ganz}) and (\ref{gmtr}) 
\begin{equation}\label{gg} 
 gg^{rs}=\frac{\delta_{rs}}{q_{r}^{2}}\prod_{t=1}^{p}q_{t}^{2},
 \ \ \
\Delta^{(p)}\vec x=-\vec x,
\end{equation}
where  $\Delta^{(p)}:=\sum_{r=1}^{p}\partial_{r}^{2}$ is 
the Laplace operator. The use of (\ref{gg}) transforms the 
system (\ref{xeqvg}), equivalent to the  Hamiltonian  
equations (\ref{hameq}), into the algorithmic chain of $p$ 
nonlinear equations for the ${\bf q}(t)$ components $q_{1},q_{2},\ldots,q_{p}$ 
\begin{equation}\label{qeqv} 
\ddot{q_{r}} + (\frac{T}{C})^2 (q_{1}\ldots 
q_{r-1} q_{r+1}\ldots q_{p})^{2} q_{r} =0, 
\end{equation}
where the component index $r$ runs from 1 to p.

Multiplication of the $r$-th equation of the system (\ref{qeqv})
 by $ q_{r}$ and subsequent summing over $r$ results in the first
 integral of Eqs. (\ref{qeqv})
\begin{equation}\label{1sti}
\dot{\bf{q}}^2+ (\frac{T}{C})^2
(q_{1} q_{2}...q_{p})^{2} = c
\end{equation}
which coincides with the initial data (\ref{gHC}) if 
the integration constant $c=1$.
 
It is easily seen that Eqs. (\ref{qeqv}) may be presented
 in a condenced form as
\begin{equation}\label{veqv} 
C\ddot{\bf{q}}=- \frac{\partial V}{\partial\bf{q}}, 
\end{equation}
where the elastic energy density  $V(\bf{q})$ turns out to be proportional 
to the determinant $g$ of the ${\cal U}$ invariant (hyper)surface of p-brane
\begin{equation}\label{poten} 
V({\bf q})=\frac{T^2}{2C} g \equiv
\frac{T^2}{2C}( q_{1}...q_{p})^{2}. 
\end{equation}
Thus, we see that the dynamics of a toroidal $p$-brane is handled by 
anharmonic elastic potential (\ref{poten}) positively defined for $C>0$.

\section{$U(1)\times U(1)\times \ldots \times U(1)$ 
invariant $p$-branes and $p$-dimensional anharmonic oscillators}

The derived equations (\ref{qeqv}) may be generated by a Hamiltonian $H$ free 
from the square root present in $H_{0}$ (\ref{sqrham}). 
Such a possibility is a consequence of our fixing the residual 
gauge symmetry  that reduces  $\mathcal{P}_0$ to the constant $C$.  
 
As a result,  $C$ can be used to write the Hamiltonian density 
 free  from square root
 $$
\mathcal{H}=\frac{\mathcal{H}_{0}^{2}}{2C}=C/2.
$$ 
Hamiltonian $H$ and PB's associated with the density $\mathcal{H}$ are 
\begin{eqnarray}
H:=\int d^p\sigma \mathcal{H}, \ \ \ 
\mathcal{H}=\frac{1}{2C} (\boldsymbol{\pi}^2 + T^{2} \prod_{a=1}^{p}q_{a}^{2}), 
\label{nosqrham} 
\\
 \{\pi_{a}, q_b\}=\delta_{ab}, \ \ \  \{q_{a}, q_b \}=0,  \ \ \ 
 \{\pi_{a}, \pi{_b} \}=0. \nonumber
\end{eqnarray}
 The Hamiltonian (\ref{nosqrham}) generates Eqs. (\ref{qeqv}) and describes
 $p$-dimensional anharmonic
 oscillator without quadratic terms associated with the harmonic oscillator. 
 The set of Hamiltonians (\ref{nosqrham}) contains potential energy 
quartic in  $\bf{q}$ for $p=2$ or presented by higher degree monomials for $p>2$. 
 
 A physical interpretation of nonlinear system  (\ref{qeqv}) 
 in terms of harmonic oscillator shows that each of the coordinates 
 $q_{r}(t)$, undergoing  the force $F_{r}(t)$,
 evolves with the instant cyclic "frequency" $\omega_r(t)$ proportional to the 
 product of all the remaining coordinates at any fixed moment $t$
\begin{equation}\label{omegi} 
 \omega_r(t)= \frac{T}{C} |q_{1}\ldots 
q_{r-1} q_{r+1}\ldots q_{p}|,\ \ \ \ r=(1,2, \ldots,  p),
\end{equation}
as it follows from (\ref{qeqv}). 
These "frequencies" $\omega_r$ cannot be infinitely large 
because of the initial data constraint (\ref{1sti}) with $c=1$
\begin{equation}\label{indat}
\sqrt{1- \dot{\bf{q}}^2}= \frac{T}{C}|q_{1} q_{2}...q_{p}|,
\end{equation}
which manifests the conservation of the energy density and shows that 
\begin{equation}\label{restr}
0 \leq |\dot{\bf{q}}| \leq 1,  \ \ \  0 \leq \frac{T}{C}|q_{1} q_{2}...q_{p}|  \leq 1.
\end{equation}
Eqs. (\ref{restr}) means that the velocity $|\dot{\bf{q}}|$ grows 
if the $p$-brane (hyper)volume $\sim |q_{1} q_{2}...q_{p}|$ diminishes 
and reaches the velocity of light ($|\dot{\bf{q}}|=1)$ while the (hyper)volume 
vanishes. The minimal velocity $\dot{\bf{q}}=0$ corresponds to the maximal 
(hyper)volume $\sim |q_{1} q_{2}...q_{p}|$ equal to $C/T$.
This  explains the finiteness of $\omega_r$ 
\begin{equation}\label{velos}
\omega_r(t)=\frac{\sqrt{1- \dot{\bf{q}}^2}}{|q_{r}|}.
\end{equation} 

In the case of tensionless p-branes \cite{Z_0, BZ_0} Eqs. (\ref{qeqv}) 
reduce to the equations
\begin{equation}\label{particl}
T=0: \ \ \  \Rightarrow \ \ \    \ddot{\bf{q}}=0,  \ \ \  |\dot{\bf{q}}| =1,
\end{equation} 
similar to the integrable equation of free massless particle in the 
effective $p$-dimensional $\bf{q}$-space. 

In general,  Eqs. (\ref{qeqv}) are  complicated owing to a monomial 
entanglement of $q_{r}$ in potential $V$ (\ref{nosqrham}). 
However, there is a solvable case discussed below.

\section{Elliptic and hyperelliptic solutions}

In the degenerate case characterized by the coincidence of all 
the components  $q_{1}= q_{2}=...=q_{p}\equiv q$ 
Eqs.  (\ref{qeqv}) are reduced to one nonlinear differential equation 
\begin{equation}\label{eleq}
\ddot{q} + (\frac{T}{C})^2 q^{(2p-1)} = 0
\end{equation} 
equivalent to its first integral (\ref{indat}) expressing the energy conservation
\begin{equation}\label{hyper} 
p\dot{q}^{2} + (\frac{T}{C})^{2}q^{2p} = 1.
\end{equation} 
After the  change of $q$ by the new variable $y =\Omega^{\frac{1}{p}}\sqrt{p}q$,
with $\Omega:=\frac{T}{C}p^{-\frac{p}{2}}$,  equation  (\ref{hyper}) takes the form
\begin{equation}\label{hypertr} 
(\frac{d y}{d\tilde{t}})^{2}= \frac{1}{2}(1-y^p)(1+y^p), 
\end{equation} 
where a new time variable  $\tilde{t}:\, = \sqrt{2}\Omega^{\frac{1}{p}}t$ is introduced. 

In the membrane case ($p=2$) Eq.(\ref{hypertr})
 coincides with the equation defining the Jacobi
 elliptic cosine $cn(x;k)$
\begin{equation}\label{elcndef} 
(\frac{d y}{d x})^{2}= (1-y^2)(1- k^2 + k^2y^2),
 \end{equation} 
if the elliptic modulus $k=\frac{1}{\sqrt{2}}$.
 Thus, $y(t)= cn(\sqrt{2\omega}t;\frac{1}{\sqrt{2}})$  with $2\omega= T/C$.
  
  After using the relation $q\equiv y/{\sqrt{2\omega}}$ we 
 obtain the elliptic solution for the desired coordinate $q(t)$, 
\begin{equation}\label{elcn} 
q(t)=\frac{1}{\sqrt{2\omega}}
cn(\sqrt{2\omega}(t+t_{0});\frac{1}{\sqrt{2}})
\equiv\sqrt{\frac{C}{T}}cn(\sqrt{\frac{T}{C}}(t+t_{0});\frac{1}{\sqrt{2}})
\end{equation} 
that is similar to the elliptic solution for $U(1)$ invariant 
membrane in $D=5$ earlier obtained in \cite{TZ}. 

If the initial velocity $\dot{q}(t_0)>0$ then
the solution (\ref{elcn}) describes an expanding torus which 
reaches the maximal size $q_{max}=\sqrt{\frac{C}{T}}$ at
some moment $t$ and then contracts to a point in a finite time
${\bf K}(1/\sqrt{2})\sqrt{\frac{C}{T}}$
(where ${\bf K}(1/\sqrt{2})=1.8451$) is the quarter period of the elliptic
cosine).

An explicit equation of the surface $\Sigma_{2}(t)$ of the
 contracting torus (\ref{elcn}) is
\[
 x_1^2+x_2^2+x_3^2+x_4^2=
\frac{4C}{T}cn\left(\sqrt{\frac {T}{C}}(t+t_0),\frac{1}{\sqrt{2}}\right)^{2},
\ \ \
  x_1x_4=x_2x_3.
\]

For the case $p>2$ integration of Eq.(\ref{hypertr}) results in the solution
\begin{equation}\label{solhyper} 
\tilde{t}=\pm\sqrt{2}\int\frac{dy}{\sqrt{1-y^{2p}}} + const
\end{equation} 
that contains hyperelliptic integral and defines an implicit dependence 
of $q$ on time.
Thus, the general solution of Eq. (\ref{hyper}) 
is expressed  in terms of hyperelliptic functions generalizing elliptic functions.  

The variable change $z=y^{2p}$ transforms the solution (\ref{solhyper}) into 
the integral
\begin{equation}\label{solhyp2} 
\tilde{t}-\tilde{t_0} 
= \pm\frac{1}{\sqrt{2}p}\int_{0}^{z^\frac{1}{2p}}dz 
z^{(\frac{1}{2p}- 1)}(1-z)^{-\frac{1}{2}}
\end{equation} 
similar to the integral discussed in \cite{LLm}.
The use of representation (\ref{solhyp2}) allows to find the contraction 
time $\Delta\tilde{t}_{c}$ 
of the 
degenerate $p$-torus from its maximal size defined  by the coinciding radii 
value  $q_{max}=(\frac{C}{T})^\frac{1}{p}$ to $q_{min}=0$. 

This time  turns out to be proportional to the well-known integral 
$$
\Delta\tilde{t}_{c}
=\frac{1}{\sqrt{2}p}\int_{0}^{1}dz z^{(\frac{1}{2p}- 1)}(1-z)^{\frac{1}{2}-1}= \frac{1
}{\sqrt{2}p}B(\frac{1}{2p}, \frac{1}{2})
$$
which defines the Euler beta function $B(\frac{1}{2p}, \frac{1}{2})$.
 
Coming back to the original time  $t$ and taking into account that $C=2E$ 
we find the desired contraction time 
\begin{equation}\label{period}
\Delta t_{c}\equiv\frac{1}{\sqrt{2}}\Omega^{-\frac{1}{p}}\Delta\tilde{t}_{c}
=\frac{1}{2\sqrt{p}}(\frac{2E}{T})^{\frac{1}{p}}
B(\frac{1}{2p}, \frac{1}{2})
\end{equation} 
as a function of the $p$-brane dimension $p$ and its energy density $E$.

So, the case of degenerate toroidal $p$-brane with coinciding radii 
is proved to be integrable and revealing the connection of the $p$-brane 
equations with the (hyper)elliptic and Euler beta functions.

\section{Anharmonic deformation of the Hooke law}

To add to understanding of the physics described by the ${\cal U}$ 
invariant p-branes one should analyse Eqs. (\ref{veqv}). 
As is seen  they are representable as the
 equations of motion of general elastic 
medium with the mass density $\rho$ 
\begin{equation}\label{lali} 
\rho\ddot{u}_{i}= \frac{\partial\sigma_{ik}}{\partial x_{k}},
\end{equation}
where $\ddot{u}_{i}$ and $\sigma_{ik}$ are the medium 
acceleration and its stress tensor \cite{LL}.
 This becomes apparent after using the symmetric stress tensor $\sigma_{rs}$ 
\begin{equation}\label{strst} 
\sigma_{rs}:=-p\delta_{rs}, \ \ \ \ \ \ \
p:=\frac{T^2}{2C} g \equiv\frac{T^2}{2C} \prod_{s=1}^{p}q_{s}^{2},
\end{equation}
 where  $p=V$ is an isotropic pressure per unit (hyper)area of
  p-brane (hyper)volume. 
Then Eqs. (\ref{veqv}) take the form similar to Eqs. (\ref{lali})
\begin{equation}\label{eleqv} 
C\ddot{q_{r}} =
- \frac{\partial{[\delta_{rs}T^{2}(2C)^{-1}(q_{1}q_{2} 
\ldots q_{p})^2}]}{\partial q_{s}}
\end{equation}
with the constant $C$ substituted for the mass density $\rho$.
   The constant $C$ is the energy density  
$\mathcal{P}_0=T|q_{1}q_{2} \ldots q_{p}|/\sqrt{1- \dot{\bf{q}}^2}$  
(\ref{gHC}) of the toroidal $p$-brane taken in the fixed gauge for the 
worldvolume diffeomorphisms. 
From physical point of view the replacement $\rho \rightarrow \mathcal{P}_0$ 
is typical of the 
standard transition to relativistic mechanics, e.g. when the rest 
energy $m$ of a non-relativistic particle is replaced by its 
relativistic energy, i.e. $m \rightarrow m/\sqrt{1- \dot{{\bf x}}^2}$ 
(where the velocity of light is absent in accordance with our agreements).

However, Eqs. (\ref{eleqv}) have no explicit relativistic covariant form 
in contrast to the original equations (\ref{5}). 
This breakdown is a result of the choice of both the gauge fixing 
and the anzats (\ref{ganz}). Such a non-relativistic formulation of the 
toroidal $p$-brane dynamics gives a reason to raise the question 
about the possibility of the existence of non-relativistic elastic media 
described by equations (\ref{eleqv}) in solid state physics. 
To answer this question let us note that    
the $p$-brane medium pressure $p$ (\ref{strst}) is created by the 
internal force $F_r$ 
\begin{equation}\label{force} 
F_{r}(t)= - \frac{\partial V}{\partial q_{r}}\equiv-\frac{T^2}{C} 
(q_{1}\ldots q_{r-1} q_{r+1}\ldots q_{p})^{2} q_{r} 
\end{equation}
 produced by the elastic potential $V$ (\ref{poten}). 
Relations (\ref{force}) manifest the breakdown of
 linear Hooke's law  and its replacement by the nonlinear law.
 They may be presented in a vector-like form 
\begin{equation}\label{force'} 
{\bf F}(t)= - \frac{\partial V}{\partial {\bf q}}\equiv-\frac{T^2}{C} 
|{\bf q}|^{2(p-1)} {\bf q} 
\end{equation}
for the case of integrable degenerate toroidal $p$-brane, 
discussed in the previous section.
Similar anharmonic generalization of the 
 harmonic Hooke law was earlier revealed in liquid crystals such as 
2d and 3d smectics $\mathcal{A}$, resulting in the 
study of elastic critical points, anisotropic scaling behaviour 
and renormalizations of 3d smectic elastic constants. 
 
So, studying the $branes \leftrightarrow smectics$ correspondence
induced by the anharmonic elasticity, characterizing both toroidal 
$p$-branes and liquid crystals, may give raise to a new insight in the 
brane physics. The well-known general intertwines of the statistical 
mechanics and field theory may also hint in behalf of such a studying.

\section{Discussion}

 Proposed is a new $U(1)\times U(1)\times\ldots \times U(1)$ invariant anzats 
describing the $p$-brane
hypersurfaces $\Sigma_{p}$ embedded in the $D=(2p+1)$-dimensional 
Minkowski space.  
The compact hypersurface $\Sigma_{p}$ is shown to be isometric 
to a flat $p$-dimensional torus with time-dependent radii.
The equations of toroidal p-brane are derived and reduced to 
an algorithmic chain of entangled equations associated 
with $p$-dimensional anharmonic oscillator. 
 Constructed is the Hamiltonian of $p$-dimensional 
anharmonic elacticity described by  the monomial potential of the 
degree $2p$  (with $p=2,3,...,(D-1)/2$ ). 
 These equations are proved to be integrable in the case of the 
degenerate $p$-torus characterized by equal radii. 
The obtained general solutions describe contracting $p$-tori and are 
presented by the 
elliptic cosine for $p$=2 or hyperelliptic functions for higher $p>2$.
The exact formula for the contraction time of the $p$-branes is found.
These results provide an example of admissible anharmonic
 potentials, associated with branes 
and, in particular, with the 5-brane of $D=11$ M/string theory.
They also demonstrate a type of possible complications accompanying 
the transition from strings to branes. 
As a result of the complications, the space of trigonometric functions 
associated with the string dynamics undergoes deformation 
controlled, in particular, by the values of the moduli of 
(hyper)elliptic functions. 

On the other hand, taking into account the geometric consideration  
of strings and branes as (hyper)surfaces embedded in the Minkowski 
space-time one may  try to construct the general solution of  
brane equations assuming the existence of corresponding  Backlund 
transformations. 
Then the application of such conceivable transformations to the 
(hyper)elliptic solutions would generate an infinite number 
of conservation laws by analogy with the case
 of strings or two-dimensional integrable sigma models \cite{Pohl}. 
This assumption is substantiated by the Regge-Lund 
geometric approach originally applied to strings \cite{RL}. 
In the geometric approach the string dynamics 
is reformulated into the consideration of the string worldsheet 
as a two-dimensional surface embedded in the four-dimensional 
Minkowski space-time. 
The projection of the string worldsheet on
a three-dimensional hypersurface of a constant time studied in \cite{RL} 
 resulted in the classical problem of surfaces as manifolds embedded 
in the three-dimensional Euclidean space. 
The Gauss-Codazzi equations fixing  the
conditions for the embedding were reduced there to the sine-Gordon equation. 
Further, the Gauss-Weingarten equations  
(the first order linear differential equations for the 
tangent and normal vectors) were proved to be presented in the form of 
the linear equations just associated with the sine-Godon equation in the 
inverse scattering method \cite{AKNS}, \cite{TF}.  
Similar results follow from the generalization of the Regge-Lund approach 
for strings evolving in higher dimensional Minkowski spaces \cite{Om},
\cite{BNC}, \cite{BN}, \cite{Z1}.
The Maurer-Cartan structure equations were used in \cite{Z1}
 to describe the string worldsheet in higher dimensional spaces.
These equations play the role of integrability 
conditions for the linear equations describing a repere  moving 
 along the string worldsheet embedded into $D$-dimensional 
Minkowski space. As a result, the Nambu-Goto string was shown  to
 describe a closed sector of the two-dimensional $SO(1,1)\times SO(D-2)$
 gauge model with the Maurer-Cartan equations reduced to the nonlinear 
chain \cite{Z1} of  PDE's  for the  $(D-2)$ sectorial curvatures 
of the string worldsheet.
This approach may be applied for the consideration of $p$-branes 
as embedded (hyper)surfaces.
In the case of $p$-brane embedded in $D$-dimensional Minkowski space-time 
 the string local group $SO(1,1)\times SO(D-2)$ has to be substituted by the 
local $SO(1,p)\times SO(D-p-1)$ group attached to the $p$-brane worldvolume.
Then the  time-dependent (hyper)elliptic solutions considered as particular 
solutions in associated  $SO(1,D-1)/SO(1,p)\times SO(D-p-1)$ sigma models 
would be used for the construction of desired general solution with help of above 
discussed Backlund transformations.

Studying the physics associated with toroidal $p$-branes 
we find that they may be treated as elastic media described by 
anharmonic generalizations of Hooke's law and subjected to
time-dependent isotropic pressure. 
As mentioned in the previous section similar type of anharmonic 
elasticity was earlier discovered in smectics $\mathcal{A}$, and therefore 
these liquid crystals can be considered as media modelling some properies 
of toroidal $p$-branes. Then one can try to apply the physics of smectics 
to the branes of M-theory to get more information on the brane  physics.

Another issue which  deserves an attention is the study of toroidal $p$-branes 
 in curved space-time and, in particular, in Anti de Sitter space-time 
for further verification of the AdS/CFT duality conjection.

\noindent{\bf Acknowledgments}

I am grateful to J. Hoppe, O. Manyuhina, A. Morozov, V. Nesterenko, K. Stelle,
 M. Trzetrzelewski, D. Uvarov and H. von Zur-Muhlen for useful discussions, 
and to Physics Department of Stockholm University, Nordic Institute for Theoretical 
Physics Nordita for kind hospitality and support of this research.


\begin{thebibliography}{99}


\bibitem{M1} P. K. Townsend, \textit{The eleven-dimensional supermembrane revisited},
Phys. Lett. B350 (1995) 184;
E. Witten, \textit{ String Theory Dynamics In Various Dimensions}, Nucl. Phys. B443 (1995) 85;
T. Banks, W. Fishler, S.H. Shenker, L. Susskind,  \textit{ M theory as a Matrix Mode:
 A conjecture}
Phys. Rev. D55(8):5112;
M. J. Duff, \textit{The world in Eleven Dimensions: Supergravity, Supermembranes and M-theory}
 (IOP, Bristol 1999).

\bibitem{Reb}
C. Rebbi, \textit{Dual models and relativistic strings}, Phys. Rep.  12 (1974) 1.

\bibitem{tucker}
R. W. Tucker, \textit{Extended Particles and the Exterior Calculus}, 
Lectures given at the Rutherford Laboratory, Feb 1976.

\bibitem{hoppe1}
J. Hoppe,
MIT PhD Thesis,1982, and in Proc. Int. Workshop on Constraints Theory and 
Relativistic Dynamics. G. Longhi, L. Lusanna (eds), Singapore:
 World Scientific 1987.

\bibitem{BST}
E. Berqshoeff, E. Sezgin, P.K. Townsend,
Phys. Lett. B189 (1987) 75.

\bibitem{DHIS}
M. Duff, P. Howe, T. Inami, K. Stelle,
Phys. Lett. B191 (1987) 70.

\bibitem{FI}
E. Floratos, J. Illipoulos,
Phys. Lett. B201 (1988) 237.

\bibitem{WHN}
B. de Witt, J. Hoppe, G. Nicolai,  Nucl.  Phys. B305 [FS23] (1988) 545.

\bibitem{Z_0}
A.A. Zheltukhin, 
 Sov. J. Nucl. Phys. 48 (1988) 375. 

\bibitem{WLN}
B. de Witt, M. Lusher, G. Nicolai,  Nucl.  Phys. B320 (1989) 135.


\bibitem{BZ_0}
I.A. Bandos and A.A. Zheltukhin,
 Fortsch. Phys. 41 (1993) 619. 

\bibitem{hoppe2}
M. Bordemann, J. Hoppe,
Phys. Lett. B317 (1993) 315
;
ibid. B325 (1994) 359. 

\bibitem{Pol}
J. Polchinski, Phys. Rev. Lett. 75 (1995) 4729.

\bibitem{UZ}
 D.V. Uvarov and A.A. Zheltukhin,
 Phys. Lett. B545 (2002) 183.

\bibitem{hoppe6}
J. Arnlind, J. Hoppe, S. Theisen,
 Phys. Lett. B 599 (2004) 118. 

\bibitem{LLm}
L.D. Landau and E.M. Lifshits,  \textit{Mechanics}, Moscow, "Science", 1965.

\bibitem{FT}
L.D. Faddeev, L.A. Takhtajan, \textit{Hamiltonian Methods in the Theory of Solitons},
"Addison-Wesley", 1987.
\bibitem{JU1}
J. Hoppe,
Compl. Anal. Oper. Theory 3 (2009) 419.
 
\bibitem{TZ}
M. Trzetrzelewski and A.A. Zheltukhin,
 Phys. Lett. B679 (2009) 523. 

\bibitem{ZT}
 A.A. Zheltukhin and M. Trzetrzelewski,
 J. Math. Phys. 51 (2010) 062303.

\bibitem{C}
P.G. Calle, C.R. Acad. Sci. Paris B274 (1972) 891.

\bibitem{GP}
G. Grinshtein and R.A. Pelcovits,
 Phys. Rev. Lett. 47 (1981) 856.

\bibitem{GZ}
L. Golubovic and Zhen-Gang Wang,
 Phys. Rev. Lett. 69 (1992) 2537.

\bibitem{LL}
L.D. Landau and E.M. Lifshits,  \textit{Theory of Elasticity},  Moscow, "Science", 1987.

\bibitem{Pohl}
K. Pohlmeyer,
Comm. Math. Phys. 46 (1976) 207. 

\bibitem{RL}
F. Lund and T. Regge,
 Phys. Rev. D 14 (1976) 1524.
 
\bibitem{AKNS}
M. Ablowitz, D.J. Kaup, A.C. Newell and H. Segur, 
 Phys. Rev. Lett. 30, (1973) 1262.

\bibitem{TF}
 L.A. Takhtajan and  L.D. Faddeev,
 Theor. Math. Phys. 21 (1974) 1046.

\bibitem{Om}
R. Omnes, Nucl. Phys. B149 (1979) 269.

\bibitem{BNC}
B.M. Barbashov, V.V. Nesterenko and A.M. Chervyakov,
 Theor. Math. Phys. 40 (1979) 15.

\bibitem{BN}
B.M. Barbashov, V.V. Nesterenko,
\textit{Introduction to the Relativistic String Theory}, World Scientific Pub Co Inc, 1990. 

\bibitem{Z1}
 A.A. Zheltukhin, Sov. J. Nucl. Phys. 33 (1979) 1723; Phys. Lett. 116B (1982) 147;
 Theor. Math. Phys. 56 (1983) 785.   



\end{thebibliography}
\end{document}